\documentclass[a4paper,11pt]{article}
\usepackage[utf8]{inputenc}
\usepackage{amsmath}
\usepackage{amsfonts}
\usepackage{amssymb}
\usepackage[margin=1.0in]{geometry}
\usepackage{graphicx}
\usepackage[swedish,english]{babel,varioref}
\usepackage{csquotes}
\usepackage[font=small, labelfont=bf]{caption}
\usepackage{subcaption}
\usepackage{multimedia}
\usepackage{import}
\usepackage{gensymb}
\usepackage{authblk}
\usepackage{bm}
\usepackage{listings}
\usepackage{xcolor}
\usepackage{tikz}
\usepackage{grffile}    
\usepackage{fancyhdr}
\usepackage{placeins}
\usepackage{array}
\usepackage{algpseudocode}
\usepackage[]{algorithm}
\usepackage{algpseudocode}
\usepackage{multirow}
\usepackage{booktabs}
\usepackage{authblk}

\usepackage[
  backend=biber,
  style=numeric-comp,
  sorting=none]
{biblatex}
\DeclareMathOperator*{\argmin}{argmin} 
\newcommand{\dtoprule}{\specialrule{1pt}{0pt}{0.4pt}%
            \specialrule{0.3pt}{0pt}{\belowrulesep}%
            }
\newcommand{\dbottomrule}{\specialrule{0.3pt}{0pt}{0.4pt}%
            \specialrule{1pt}{0pt}{\belowrulesep}%
            }

  {\par\vspace{\abovedisplayskip}\noindent\begin{tabular}{>{$}l<{$} @{${}={}$} l}}
  {\end{tabular}\par\vspace{\belowdisplayskip}}

\usetikzlibrary{external,positioning} 

\title{Optimizing wheel loader performance --- an end-to-end approach}

\author[1,2]{Koji Aoshima}
\author[3]{Eddie Wadbro}
\author[1]{Martin Servin \thanks{martin.servin@umu.se}}

\affil[1]{Umeå University}
\affil[2]{Komatsu Ltd}
\affil[3]{Karlstad University}

\date{}

\begin{document}

\maketitle

\begin{abstract}
Wheel loaders in mines and construction sites repeatedly load soil from a pile to load receivers. 
  Automating this task presents a challenging planning problem since each loading's performance depends on the pile state, which depends on previous loadings.
  We investigate an end-to-end optimization approach considering future loading outcomes and transportation costs between the pile and load receivers.
  To predict the evolution of the pile state and the loading performance, we use
  world models that leverage deep neural networks trained on numerous simulated loading cycles.
  A look-ahead tree search optimizes the sequence of loading actions by evaluating the performance of thousands of action candidates, which expand into subsequent action candidates under the predicted pile states recursively.
  Test results demonstrate that, over a horizon of 15 sequential loadings, the look-ahead tree search is 6\,\% more efficient than a greedy strategy, which always selects the action that maximizes the current single loading performance, and 14\,\% more efficient than using a fixed loading controller optimized for the nominal case.
\end{abstract}

\section{Introduction}
The task of most construction and mining machines, such as wheel loaders, is to perform long sequences of earthmoving operations in dynamic environments. 
Optimizing this task involves selecting loading and transportation actions that maximize the gain over time, according to some performance measure.
This is challenging for several reasons. First, the performance of a single loading cycle depends highly on the local shape of the pile and the soil properties \cite{singh2006factors}. 
The loading actions need to be carefully adapted to these conditions. 
Second, each loading alters the pile shape. 
From the perspective of future loadings, such a change can be either for the better or worse. 
There might be actions that are of low performance in the short term but are necessary to achieve higher gains in the future \cite{singh1992task}.
Therefore, a greedy strategy of always choosing the loading action that maximizes the performance for a single loading might be sub-optimal in the long run. 
Such a strategy might gradually deteriorate the pile state or lead to excessive transportation between the dig location and the load receiver.
This motivates considering future states when selecting the actions.


We refer to the problem of finding the action sequence that maximizes the total performance of many wheel loading cycles, given the initial pile state, as the \emph{wheel loader end-to-end optimization problem}.
To the best of our knowledge, this problem has not been scientifically explored previously.
In fact, until recently, it was computationally intractable.
In our previous work \cite{aoshima2023predictor}, we developed data-driven models capable of predicting the loading performance and the subsequent state of the pile within a few milliseconds.  Model inputs are the previous pile state and selected action parameters.
This makes it possible to rapidly evaluate the net performance of numerous alternative action sequences in advance.
These models are referred to as a \emph{world model}, as they predict the new state the environment will transition into, given the previous state and a selected action, and the observations associated with the state transition \cite{ding2024}.

This paper investigates methods for solving the wheel loader end-to-end optimization problem and analyzes the obtained solutions.
The research questions include what characterizes the optimal action sequence, how sensitive the results are to the initial pile state and planning horizon, what the computational demands are, and the feasibility of practical use of the method. The idea is that the methods could be implemented in site management systems with high-precision monitoring of equipment and the 3D topography, sending coordinated work plans to the individual machines.

\begin{figure}[!htb]
    \centering
    \includegraphics[width=0.9\textwidth]{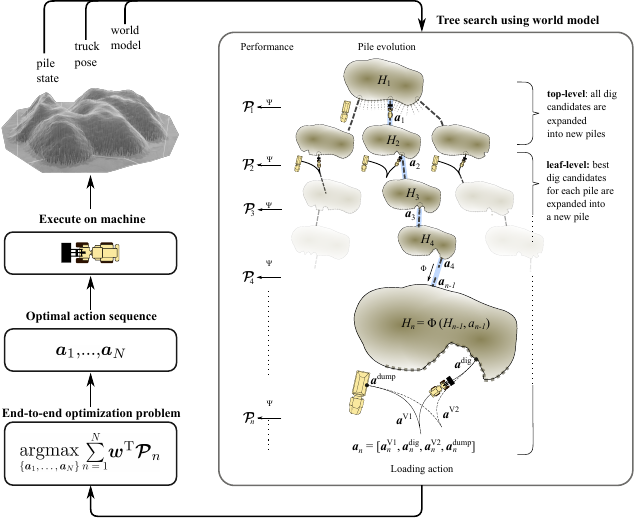}
    \caption{Overview of the look-ahead tree search algorithm for a wheel loader moving soil in V-shaped loading cycles from the evolving pile to a dump truck.}
    \label{fig:overview}
\end{figure}

We use a look-ahead tree search approach illustrated in Fig.~\ref{fig:overview}.
Future action candidates for the foreseen future states are identified recursively, offline. The action candidates are evaluated, using data-driven 
models developed in previous research \cite{aoshima2023predictor}, to compute their 
respective performance and expand into further action candidates under the new states. 
The method should apply to any kind of world model with the same prediction capabilities.
This approach is, reportedly, similar to how experienced operators intuitively plan their work, but we expect that a computerized model 
and search algorithms can make more accurate predictions and plan over longer horizons.
To understand the importance of the look-ahead search, we examined the effect of using different planning horizons and objectives.

The wheel loader is assumed to have a high-level planner and a low-level control system, which 
work as the hierarchical architecture described in refs. \cite{lever2001} and \cite{wang2021}.
The high-level planner selects a subtask action, such as bucket filling, dumping, or V-shaped transportation, according 
to the current state of the machine and the work site. 
A low-level controller performs the subtasks given a motion plan and control parameters.
For bucket filling, the machine is equipped with a type of admittance controller \cite{Dobson2017}.
Our controller takes four action parameters that determine how reactive the boom and bucket actuation is in response to the momentaneous dig force \cite{aoshima2023predictor}. 
The performance of a loading cycle is measured in terms of mass, time, and work, with
contributions from each subtask.

We test the effect of the look-ahead tree search by considering a wheel loader tasked with performing 15 sequential loading cycles.
The resulting performance and computational cost are compared with that obtained with a greedy search and a nominal strategy.

\section{Related work} 
\label{sec:related_work}
The end-to-end optimization problem of sequential loading was discussed in \cite{singh1992task}, considered intractable, and substituted with a simpler problem of optimizing each loading cycle by searching in a lower-dimensional action space implicitly constrained by the pile state.
An average of 90\,\% efficiency (bucket payload relative to its capacity) was achieved over 20 cycles. 
The study was limited to 2D, used a heuristic simulation model for soil displacements and settling, and did not consider
time or energy consumption. 
Attempts were made to improve the performance further by using heuristics to avoid action plans that 
produce challenging terrain configurations, but the optimization was computationally prohibitive. 

What is a good dig location in the pile, given its shape in terms of a heightmap, has been studied from the perspective of both bucket filling 
and the transportation between the dig and the dump location. In \cite{singh1998multi}, a coarse-level planner
finds the points along the pile contour closest to the load receiver, and a refined planner was used to avoid curved regions of the pile surface to minimize bucket-side loading and drop-in filling efficiency. 
A sequential coarse-to-fine planner was studied in \cite{magnusson2015quantitative}
and compared to a greedy planner. The coarse planner first produces a sequence of dig regions ordered radially around the pile centroid.
The fine planner then selects the dig points inside each region that have the best local pile shape.
Simulations (cellular automata) of 50 consecutive loadings showed that the coarse-to-fine strategy
maintains a good pile shape and performs at 80--90\,\% while a greedy strategy (always selecting the dig location with optimal pile shape) initially performs equal or better for ten loadings but drops down to about 60\,\% after 25 loadings. 

Models for adapting the loading control to the pile shape have been studied with different approaches.
In \cite{backman2021continuous}, reinforcement learning was used to train a multiagent system for autonomous control of an underground
load-haul-dump (LHD) vehicle using a depth camera, lidar, and force and kinematics sensors. 
One agent was used to select a favorable dig position, given the observed pile shape.
A second agent had the task of steering the vehicle toward the target position and controlling 
the forward drive as well as the lift and tilt cylinders to perform bucket filling.
The agents were trained in a simulated environment to learn control policies designed for high productivity and energy efficiency over multiple loading cycles while also avoiding collisions and wheel slip. 
The system achieved bucket filling with 75\,\% of maximum capacity on average and used 4\,\% less
energy by actively selecting favorable dig locations. The vehicle was confined to operate in a narrow mine drift with little variability in viable dig position or how to navigate to and from the dump location. 
It is untested how well this approach applies to controlling a wheel loader above ground with much larger action and observation space. The study \cite{lindmark2018computational} explored multiobjective optimization of an LHD control strategy using surrogate models based on data from sequences of simulated loadings in a pile of fragmented rock. 
Each simulated sequence used a parameterized dig trajectory planner that adapts to the local pile state. 
Based on many simulations with different parameters, the learned model can identify the loading strategies that ensure high average performance and avoid those that systematically deteriorate the pile state.

Optimization of the V-shaped transportation path between the dig and dump location has been addressed in several studies, focusing on finding the shortest path with curvature consistent with the vehicle's minimum turning radius \cite{Sarata2005, takei2015simultaneous}, 
taking into account the vehicle dynamics and construction working site constraints
\cite{ALSHAER20135315}, non-uniform trajectories under forward and
reverse driving conditions \cite{SHI2020103570}, or minimizing fuel consumption \cite{hong2017path}.
Time and energy-efficient tracking of a planned path has been explored in several studies using optimal control \cite{NEZHADALI20161,sardarmehni2023path}.


The optimal bucket filling trajectory was researched in \cite{filla2017towards} using the discrete element method (DEM) and used in \cite{frank2018optimal} to find the optimal short loading cycle using dynamic programming. 
For the sake of computational efficiency, the simulations were limited to quasi-2D piles sloped with the material's angle of repose.
The effect of the pile shape on the optimal trajectory or the resulting 3D pile state has not been explored using DEM simulation, apart from the work in \cite{lindmark2018computational}.


\section{Problem formulation} 
We focus on the \emph{short loading cycle}, in which the wheel loader repeatedly loads and dumps soil from a pile to a load receiver. 
Each cycle can be divided into a sequence of subtasks: V-turn-1, loading, V-turn-2, and dumping. These subtasks are illustrated in Fig.~\ref{fig:overview}.
We assume the wheel loader is equipped with a low-level control system for each subtask. 
A high-level system (operator or agent) activates the low-level systems by selecting a set of \emph{subtask 
action parameters}: $\bm{a}^\text{V1}$, $\bm{a}^\text{load}$, $\bm{a}^\text{V2}$, and $\bm{a}^\text{dump}$, respectively.  

At the beginning of each cycle $n=1,2,\hdots,N$, the wheel loader is located by the load receiver at a position $\bm{x}_n^\text{dump}$.
The V-turn-1 subtask is to drive the wheel loader to a selected loading position $\bm{x}_n^\text{dig}$ somewhere
along the edge of the nearby pile. A low-level controller plans and executes a V-turn motion that starts at $\bm{x}_n^\text{dump}$
and ends at $\bm{x}_n^\text{dig}$, avoiding collisions with the environment, and with endpoints heading normal to the load receiver
and the pile, respectively.  As subtask action parameters, we use the target loading position, 
$\bm{a}_n^\text{V1} = \bm{x}_n^\text{dig}$.

The next subtask is the actual loading from the pile. It starts with the machine approaching the selected loading
position and ends with the vehicle reversing from the pile with a filled bucket. The task is carried out by an automatic bucket filling controller with some 
action parameters, $\bm{a}_n^\text{load}$, that may be adapted to the current pile shape and material properties.    
The loading transforms the pile from a state $\bm{H}_n$ into a new state $\bm{H}_{n+1}$
and results in some amount of soil mass $M_n$ in the bucket. We identify the state with the geometric shape of the pile.

After bucket filling, the V-turn-2 subtask is performed. It ends with the wheel loader approaching the receiver located at
a position $\bm{x}_n^\text{dump}$, which is the corresponding action parameter $\bm{a}_n^\text{V2}$.
The final subtask is to empty the bucket's contents onto the load receiver with an action $\bm{a}_n^\text{dump}$. When the receiver is filled,
this subtask might require some adjustment action to prevent spillage.
A completed loading cycle thus involves a selection of action parameters that we collect in a vector
$\bm{a}_n = [\bm{a}_n^\text{V1}, \bm{a}_n^\text{load}, \bm{a}_n^\text{V2}, \bm{a}_n^\text{dump}]$.

A loading cycle involves spending some amount of mechanical work $W_n$ and time $T_n$ to displace a certain mass $M_n$ from the pile to the receiver. Each of the subtasks contributes to the cycle time and work, while it is only the loading subtask that produces a mass measurement. We attribute each loading cycle a performance, which we
measure by the performance vector
\begin{equation}
    \label{sec:performance_measures}
    \bm{\mathcal{P}}_{n} =  \left[\frac{M_0}{M_n}, \frac{T_n}{T_0}, \frac{W_n}{W_0} \right]^\mathsf{T},
\end{equation}
where $M_0$, $T_0$, and $W_0$ are some characteristic values used for normalization.    
Note that the performance vector is a function of the pile state, the location of the load receiver, and the selected action, i.e.,
$\bm{\mathcal{P}}_{n}(\bm{H}_n, \bm{a}_n)$.
We pose the end-to-end optimization problem of $N$ sequential loading cycles as the problem of
finding the sequence of actions $\{\bm{a}_{n}\}_{n=1}^N$ that, given the initial pile state $\bm{H}_1$ and the dump location $\bm{x}^{\text{dump}}$, satisfy 
\begin{equation}
    \argmin_{\{\bm{a}_1, \hdots, \bm{a}_N\}} \sum_{n=1}^N \bm{w}^\mathsf{T}\bm{\mathcal{P}}_{n},
\end{equation}  
where $\bm{w}$ is a vector of positive weights for controlling the trade-off between maximizing loaded mass
and minimizing time and work.

The posed optimization problem is computationally challenging to solve.
An action sequence $\{\bm{a}_{n}\}_{n=1}^N$ renders a sequence of pile states $\{\bm{H}_{n}\}_{n=1}^N$ and performance measurements $\{\bm{\mathcal{P}}_{n}\}_{n=1}^N$. These cannot be computed independently due to the strong dependency on the evolving pile state. The computational complexity for exhaustive search scales exponentially as $(t_{\bm{a}} D_{\bm{a}})^N$, where $D_{\bm{a}}$ is the number of action candidates that must be evaluated for each pile state and $t_{\bm{a}}$ is the computational time for doing so.

\subsection{Assumptions and delimitations}  
\label{sec:assumptions_and_delimitations}
A number of delimitations and simplifying assumptions are made to bring down the difficulty
of the problem.
We focus on loading a single type of non-cohesive and homogeneous soil, namely gravel.
The surrounding ground is assumed
to be flat.
We assume no soil is spilled from the bucket during the V-turn-2 subtask or when dumped on the receiver.
Spillage would affect subsequent performance, either by loss in control precision or traction 
or by the need to clear the ground.

The receiver is located at a fixed location and orientation relative to the pile.
We assume it is immediately replaced by another receiver at the same location
when full. The wheel loader always approaches the receiver's center position, orthogonally, 
and simply empties the bucket without considering the shape of the body. The contribution
to the cycle performance is then a constant value. We thus ignore the selection of the dumping action parameter $\bm{a}^\text{dump}$ and the V-turn-2 parameter $\bm{a}^\text{V2}$ from the problem but account for the contribution of the actions to the net performance.

Automatic bucket filling starts with the loader heading towards a dig location
along the edge of the pile at a nominal target speed. To limit the dimensionality of the problem, the 
heading is always normal to the pile contour, and the edge is discretized into a finite number 
of candidates of dig locations separated by a step size smaller than the width of a bucket. 

We assume there is always room for a collision-free V-turn between the receiver and the pile. As the dump location is known, the V-turn planner requires only the dig location and heading as input.
The V-turns are performed with some performance $\mathcal{P}^\text{V1}_n$ and $\mathcal{P}^\text{V2}_n$,
and we assume that we can predict them (Sec.~\ref{sec:vturn_prediction}) with sufficient accuracy given the path and the bucket load mass.

For the loading, we assume access to a world model in the following form    
\begin{align}   
    \bm{H}_{n+1} & = \bm{\Phi}    ( \bm{H}_n, \bm{a}_n ),\\
    \bm{\mathcal{P}}_n^\text{load} & = \bm{\Psi} ( \bm{H}_n, \bm{a}_n ),
\end{align}
where $\bm{\Phi}$ is a pile state predictor model
and $\bm{\Psi}$ is a performance predictor model,
which take as input the previous  pile state $\bm{H}_n$ and 
the reduced action control parameter $\bm{a}_n = [\bm{x}^\text{dig}_n,\bm{a}^\text{load}_n]$.

With these simplifications, the problem to solve is
\begin{equation}
    \label{eq:simplified_opt}
    \argmin_{\{\bm{x}^\text{dig}_n, \bm{a}^\text{load}_n\}_{n=1}^N} \sum_{n=1}^N \bm{w}^\mathsf{T}\bm{\mathcal{P}}_{n}.
\end{equation}
with the predicted performance $\bm{\mathcal{P}}_{n} = \bm{\mathcal{P}}^\text{load}_{n} + \bm{\mathcal{P}}^\text{V1}_{n} + \bm{\mathcal{P}}^\text{V2}_{n}$ that must be computed sequentially along with the predicted pile state evolution $\{\bm{H}_n\}_{n=1}^N$ starting from the initial state $\bm{H}_1$.

\section{Method} 
This section describes our models for predicting the outcome of a single loading cycle and the search method to find the optimal sequence of wheel loader actions.

\subsection{Loading prediction} 
\label{sec:loading_prediction}
We use the data-driven world model developed in \cite{aoshima2023predictor} for a Komatsu WA320-7 wheel loader doing automatic bucket filling in piles of gravel using an admittance controller.
The models predict the loading outcome
\begin{align}   
    \bm{H}' & = \bm{\Phi}    ( \bm{H}, \bm{x}^\text{dig}, \bm{a}^\text{load} ), \label{eq:pile_state_predictor}\\
    \bm{\mathcal{P}}^\text{load} & = \bm{\Psi} ( \bm{H}, \bm{x}^\text{dig}, \bm{a}^\text{load} ), \label{eq:performance_predictor}
\end{align}
in terms of the resulting heightmap $\bm{H}'$ and loading performance, $\bm{\mathcal{P}}^\text{load} = \left[M_0/M_\text{load}, T_\text{load}/T_0, W_\text{load}/W_0 \right]^\mathsf{T}$,
given the current heightmap $\bm{H}$, dig location $\bm{x}^\text{dig}$, and action parameters $\bm{a}^\text{load}$ for the automatic loading controller.

\begin{figure} [!htb]
    \centering 
    \includegraphics[width=0.75\textwidth, trim = 0 0 0 0, clip]{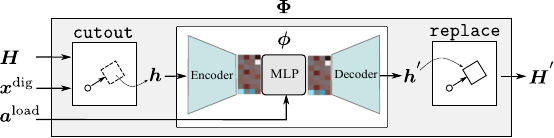}
    \caption{Illustration of the pile state predictor model.
    The neural network, $\phi$, has an encoder-decoder structure.}
    \label{fig:pile_state_prediction_overview}
\end{figure}

\begin{figure} [!htb]
    \centering 
    \includegraphics[width=0.5\textwidth, trim = 0 0 0 0, clip]{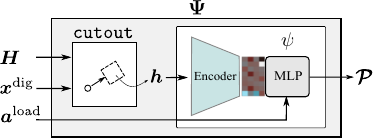}
    \caption{Illustration of the performance predictor model. 
    Since the encoder does not depend on $\bm{a}^\text{load}$, it and its gradient only need to be computed once for each dig location.}
    \label{fig:performance_prediction_overview}
\end{figure}

The dependency on the pile shape is limited to the \emph{local} pile shape, $\bm{h}$, in the area of the dig location and with the selected heading. Therefore,
the function $\bm{\Phi}$ first does a \texttt{cutout} operation of the global heightmap to obtain the local pile surface. This is then fed as input, along with the loading action parameters, to a deep neural network $\bm{\phi}$. 
A \texttt{replace} function then substitutes the predicted local heightmap $\bm{h}' = \bm{\phi}(\bm{h}, \bm{a}^\text{load})$ into $\bm{H}$ to obtain the predicted global heightmap $\bm{H}'$. 
Similarly, the loading performance is predicted using a deep neural network, $\bm{\mathcal{P}}^\text{load} = \psi(\bm{h},\bm{a}^\text{load})$, after first cutting out the local heightmap at the dig location and selected heading. 
The model has three convolutional layers to encode $\bm{h}$ before being fed with $\bm{a}^\text{load}$ to a multi-layer perceptron. 
The swish activation is used for all layers. 
The pile state and performance predictor models are illustrated in Fig.~\ref{fig:pile_state_prediction_overview} and Fig.~\ref{fig:performance_prediction_overview}, respectively. 

The models The models are trained and validated on a dataset from more than 10,000 random loading actions on various gravel pile shapes. The trained models, with roughly 10e7 parameters, achieved around 95\% accuracy in predicting loading performance and 97\% accuracy in predicting the resulting pile state. 

As in \cite{aoshima2023predictor}, we use different sizes and resolutions of the local heightmap for the pile state and performance predictor models.
For the former model, we discretize a quadratic heightmap with 5.2 m sides in a $52\times 52$ grid. For the latter model, we use a $36\times 36$ grid and a side length of 3.6 m.

\subsubsection{Optimal loading action parameters} 
The wheel loader is equipped with the function of automatic bucket filling
using admittance control. The controller is parameterized with four parameters $\bm{a}^\text{load} \in [0,1]^4$ that control how reactive 
the boom lift and bucket tilt are to the perceived digging resistance, which in turn is an unknown function of the local pile shape.
For each pile shape and soil strength, there exists some loading control parameters $\bm{a}^\text{load*}$ that optimize the loading performance, 
\begin{equation}
    \bm{a}^{\text{load}*} = \argmin_{\bm{a}^\text{load}} {\bm{w}}^\mathsf{T}\bm{\mathcal{P}}^\text{load}.
    \label{eq:optimal_loading_action}
\end{equation}
The optimal control parameters can be computed using the gradient descent method with the iterative step 
\begin{equation}
    \bm{a}^\text{load} := \bm{a}^\text{load} - \eta \nabla \left( \bm{w}^\mathsf{T} \bm{\mathcal{P}}^\text{load} \right),
    \label{eq:gradient_descent}
\end{equation}
where $\eta$ is the step length and $\nabla = \partial/\partial \bm{a}^\text{load}$. As an initial guess, we use $\bm{a}^\text{load}_n = [1,0, 0.0, 0.5, 0.5]^\mathsf{T}$, which corresponds to the machine thrusting deep into the pile, filling the bucket with little or no lifting of it \cite{aoshima2023predictor}.
This type of action has been suggested for uniformly sloped piles of dry homogeneous soil \cite{Bradley1998}, and we expect the optimization process to converge faster with this initialization.
The maximum number of iterations is set to 30 with early stopping (patience 3 for tolerance $10^{-4}$).
The gradient with respect to $\bm{a}^\text{load}$ is calculated by using \texttt{pytorch.autograd}
and exploiting that $\psi$ has two input branches, separating $\bm{h}$ and $\bm{a}^\text{load}$ so that the encoding part only needs to be evaluated once during the optimization process.


\subsection{V-turn model} 
\label{sec:vturn_prediction}
\begin{figure} [!tbh]
    \centering
    \includegraphics[width=0.40\textwidth, trim = 0 0 0 0, clip]{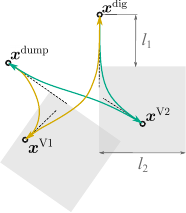}
    \caption{An example of V-turn-1 (gold) and V-turn-2 (cyan) connecting $\bm{x}^\text{dump}$ and $\bm{x}^\text{dig}$ via the switch back points $\bm{x}^\text{V1}$ and $\bm{x}^\text{V2}$ that must lie inside the square search regions (grey). Headings are indicated with dashed lines.}
    \label{fig:spline_curve}
\end{figure}

A loading cycle involves moving the wheel loader back and forth between the pile and the load receiver in a V-shaped pattern connecting the dump location $\bm{x}^\text{dump}$ and dig location $\bm{x}^\text{dig}$.
We estimate the transportation time and work by numerical integration of an assumed equation of motion for the wheel loader, including its motor strength, variable mass, and rolling resistance. 
For generating the V-turn paths, we use cubic B-splines \cite{piegl1996nurbs}.
In narrow spaces, Dubins curves may be the preferred choice to ensure that the vehicle's minimum turning radius is respected \cite{dubins1957curves}.
Each V-turn path is formed by the combination of two spline curves. V-turn-1 is composed of the splines connecting $\bm{x}^\text{dump}\to \bm{x}^\text{V1}$ and $\bm{x}^\text{V1} \to\bm{x}^\text{dig}$ with condition that the ingoing and outgoing directions should match in the switch back point $\bm{x}^\text{V1}$.
Similarly, V-turn-2 is composed of two splines connecting $\bm{x}^\text{dig}\to \bm{x}^\text{V2}$ and $\bm{x}^\text{V2} \to\bm{x}^\text{dump}$. 
See Fig.~\ref{fig:spline_curve} for an illustration.

To compute the splines, we follow the method and notations of \cite{MAEKAWA2010350}.
The path to be found is represented by five control points, $\{ \bm{p}_i\}_{i=0}^5$. These are obtained by solving the following equation system 
\begin{equation}
\null\hspace{-7pt}
    \begin{bmatrix}
        1 & 0 & 0 & 0 & 0 & 0 \\
        N_0^{1}(s_0) & N_1^{1}(s_0) & 0 & 0 & 0 & 0 \\
        N_0^{2}(s_0) & N_1^{2}(s_0) & N_2^{2}(s_0) & 0 & 0 & 0 \\
        0 & 0 & 0 & N_3^{2}(s_1) & N_4^{2}(s_1) & N_5^{2}(s_1) \\
        0 & 0 & 0 & 0 & N_4^{1}(s_1) & N_5^{1}(s_1) \\
        0 & 0 & 0 & 0 & 0 & 1
    \end{bmatrix}
    \begin{bmatrix}
        \bm{p}_0 \\
        \bm{p}_1 \\
        \bm{p}_2 \\
        \bm{p}_3 \\
        \bm{p}_4 \\
        \bm{p}_5
    \end{bmatrix}
    =
    \begin{bmatrix}
        \bm{q}_0 \\
        \alpha \bm{d}_0 \\
        \bm{0} \\
        \bm{q}_1 \\
        \beta \bm{d}_1 \\
        \bm{0}
    \end{bmatrix},
\end{equation}
where $N_i^{(j)}(s)$ is the $j$th derivative of the $i$th spline basis function associated with $\bm{p}_i$ and curve parameter $s\in[0,1]$. On the right-hand side, $\bm{q}_0$ and $\bm{q}_1$ are the spline endpoints with first derivatives $\alpha \bm{d}_0$ and $\beta \bm{d}_1$, respectively, where $\alpha$ and $\beta$ are parameters for controlling the magnitudes of the derivatives. 
For a given set of endpoints, $\bm{x}^\text{dump}$ and $\bm{x}^\text{dig}$ (and heading at these points), we compute the connecting splines, via $\bm{x}^\text{V1}$ and $\bm{x}^\text{V2}$, that minimize the path length and curvature according to the following weighted objective function \cite{MAEKAWA2010350}
\begin{equation}
    \label{eq:curve_energy_and_distance}
    \gamma_1 \sum \left( \kappa_i \right) ^2 \Delta s_i + \gamma_2 \sum \left( \cfrac{d\kappa_i}{ds} \right) ^2 \Delta s_i + \gamma_3 \sum \Delta s_i.
\end{equation}
with the weights set to $\bm{\gamma} = [10.0, 10.0, 1.0]$ to avoid sharp curves.
We use Powell's method to limit the search space such that the switch back points must reside in a box with side length $l_2 = 10$ m positioned at $l_1 = 5$ m from the endpoints, see Fig.~\ref{fig:spline_curve} for an illustration.  These constraints are also chosen to shorten the reversing distance from $\bm{x}^\text{dump}$ and $\bm{x}^\text{dig}$ to the switch back points. This explains why the V-turn-1 and V-turn-2 paths are not identical although generated using the same $\bm{x}^\text{dump}$ and $\bm{x}^\text{dig}$. 
The magnitude of the derivatives at $\bm{x}^\text{dump}$, $\bm{x}^\text{dig}$, $\bm{x}^\text{V1}$, and $\bm{x}^\text{V2}$ are fixed at 10, 30, 5, and 5, respectively, for the paths to be straight around the data points.
The implementation for computing the paths uses \texttt{scipy.optimize}. 

\begin{figure}[!htb]
    \centering
    \begin{subfigure}[b]{0.5\textwidth}
        \centering
        \includegraphics[width=1.0\textwidth, trim = 0 0 0 0, clip]{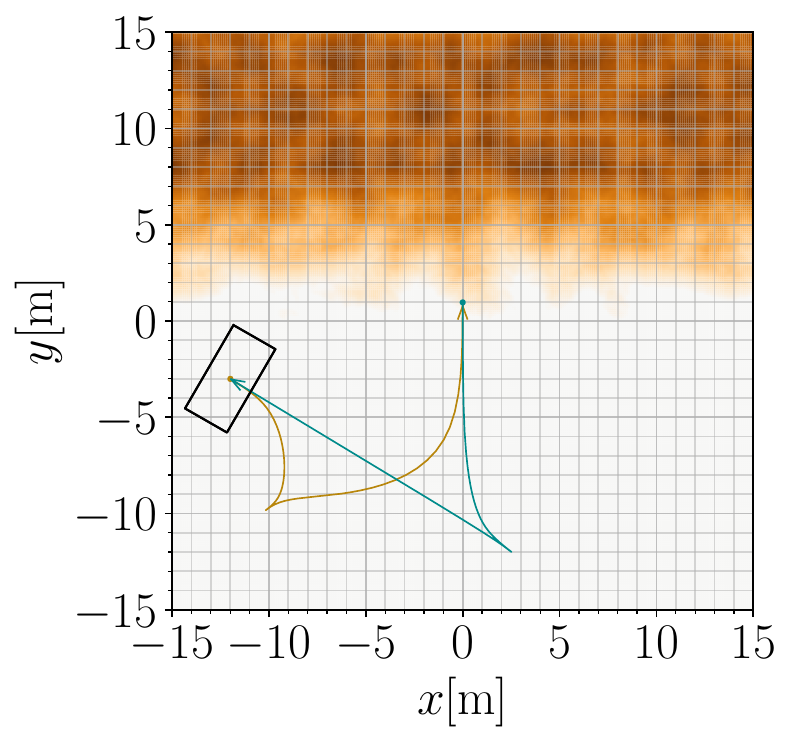}
        \caption{V-turn trajectories}
    \end{subfigure} \\
    \begin{subfigure}[b]{0.60\textwidth}
        \centering
        \includegraphics[width=1.0\textwidth, trim = 0 0 0 0, clip]{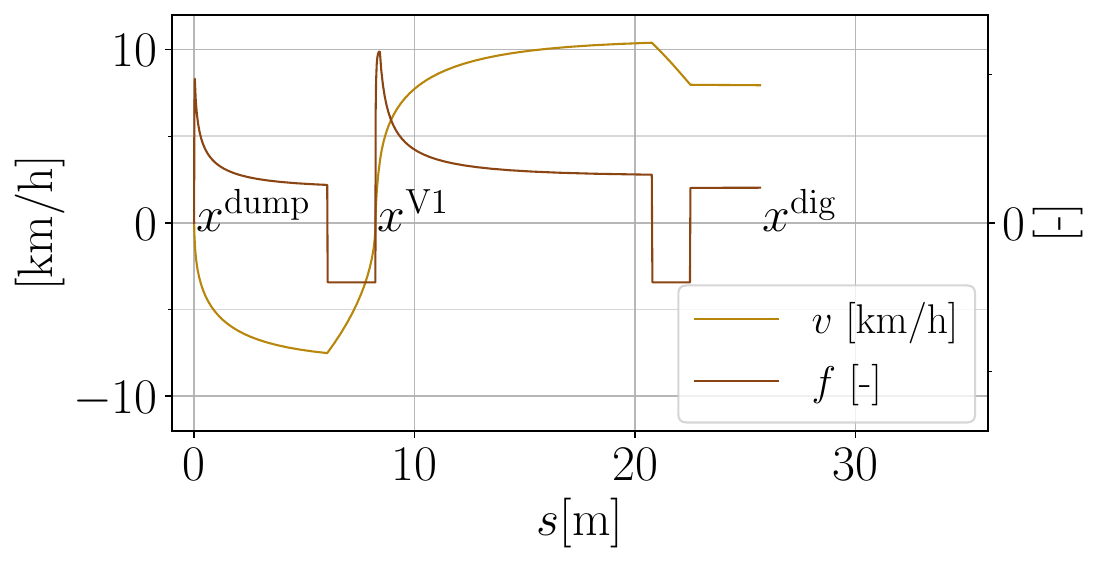}
        \caption{The speed and the force at V-turn-1}
    \end{subfigure}
    \begin{subfigure}[b]{0.60\textwidth}
        \includegraphics[width=1.0\textwidth, trim = 0 0 0 0, clip]{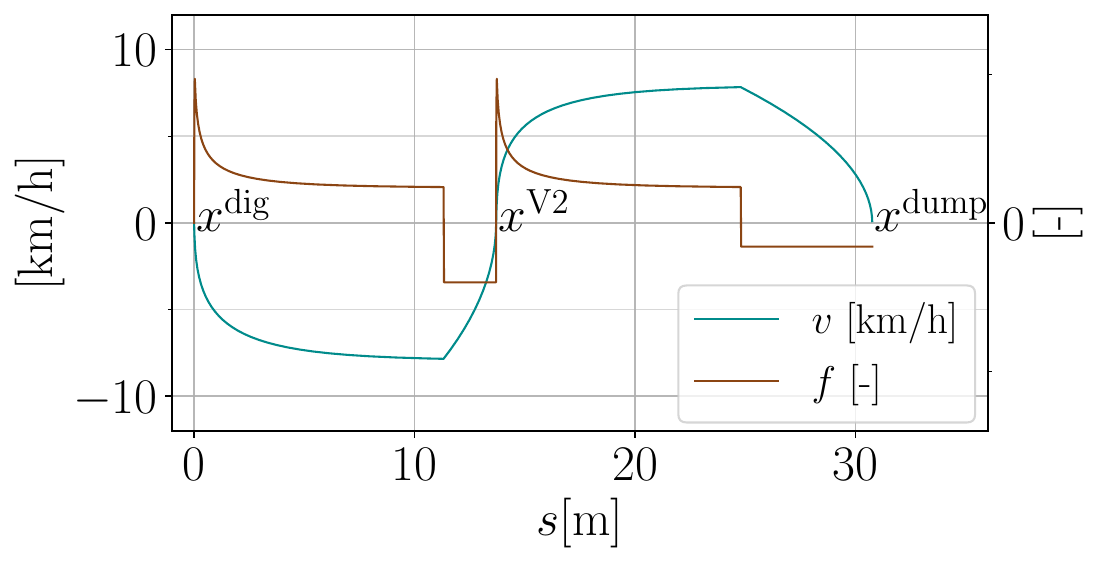}
        \caption{The speed and the force at V-turn-2}
    \end{subfigure}
    \caption{An example of the speed and the force at V-turn trajectories.}
    \label{fig:vturn_simulation_example}
\end{figure}

For each given V-turn path, the time and work for driving the vehicle is computed.
The first step is to compute the velocity profile $v(t)$ and traction force $f(t)$.
The velocity is obtained by  numerical integrations of the equation of motion
\begin{equation}
    \label{eq:acc_model}
    M_\text{tot} \cfrac{dv}{dt} + C_\text{r} v = f,
\end{equation}
where the total variable mass $M_\text{tot} = M_\text{vehicle} + M_\text{load}$ is composed of the vehicle mass $M_\text{vehicle}$, $M_\text{load}$ is the loaded mass (zero for V-turn-1), $C_\text{r}=\mu_\text{r}M_\text{tot}g$ is the rolling resistance with gravity acceleration $g$ and rolling resistance coefficient $\mu_\text{r}$.
The machine is accelerated by a traction force $f(t)$ to reach a preset target velocity during the transportation phase. We control $f(t)$ via the throttle as described in \cite{aoshima2021}. When approaching an endpoint, the machine is decelerated by a constant brake force $f=f_\text{brake}$.
The V-turn duration is computed by the time difference between the end-points and the work as $\int \text{max}[0,f(t)] v(t) dt$ along the path. 
Note that only positive work is accounted for in the work computation, i.e., no energy can be accumulated during the decelerating phase.
Also, we assume the vehicle follows the prescribed path precisely, with no required work for steering and with the bucket and boom angles held fixed during transport. 
We use a time step of 10 ms, $M_\text{vehicle} = 15,2$ tonne, $\mu_\text{roll} = 0.01$, and the input throttle value can increase with a rate of $2.0$ units/s.
The target speed is 8.0 km/h for both reversing and forwarding except for the nearest 5 m before $\bm{x}^\text{dig}$ where it increases to 11.4 km/h.
Example V-turns and trajectories with the resulting speed and traction force are shown in Fig.~\ref{fig:vturn_simulation_example}.

To save computational time during end-to-end optimization, we pre-compute
many optimal V-turns using a grid of dig locations and different headings. 
The result is stored in a look-up table. 
During optimization, the estimated time and work for the V-turns are interpolated from the look-up table.

\subsection{Dumping} 
\label{sec:dumping}
We assume the machine always empties the bucket at a fixed location at the receiver, without considering the shape of the loaded mass.
Therefore, the emptying time is fixed at 5 s, and no work is associated with the dumping action.

\subsection{Look-ahead tree search} 
\label{sec:search_method}
To find a near-optimal solution, we develop a look-ahead tree search algorithm that combines the world model and V-turn model for long-horizon predictions. The planning horizon is denoted by integer $N$ and the search depth by $d\leq N$. 
At each planning step, $n = 1,2,\hdots, N$, a finite set of $I$ candidate dig locations $\{\bm{x}_i^\text{dig}\}_{i=1}^I$ and corresponding loading actions $\{\bm{a}_i^\text{load}\}_{i=1}^I$ are considered. 
We are to select the dig location and loading action that are optimal over the search depth $d$.
At the most shallow search depth, $d =1$, this corresponds to picking the candidate that minimizes $\bm{w}^\mathsf{T}\bm{\mathcal{P}}_n(\bm{H}_n, \bm{x}^\text{dig}_n, \bm{a}^\text{load}_n)$. At the first planning step, $n = 1$,
we know the pile state $\bm{H}_1$ and can make this pick. For the future planning steps, we must first expand the pile into its next state from the previous, $\bm{H}_{n+1} = \bm{H}'_{n} \equiv \bm{\Phi}(\bm{H}_{n}, \bm{x}^\text{dig}_{n}, \bm{a}^\text{load}_{n})$.
At larger search depths, $d\geq 2$, we compute the predicted net performance over the search horizon $d$ using the following approximation of the evaluation function
\begin{align}\label{eq:Q}
    Q^d_n & = \sum_{k=n}^{n-1+d} \bm{w}^\mathsf{T}\bm{\mathcal{P}}_k(\bm{H}_k, \bm{x}^\text{dig}_k, \bm{a}^\text{load}_k) \nonumber \\
        & \lesssim \bm{w}^\mathsf{T}\bm{\mathcal{P}}_n(\bm{H}_n, \bm{x}^\text{dig}_n, \bm{a}^\text{load}_n) + \sum_{k=n+1}^{n-1+d} \bm{w}^\mathsf{T}\bm{\mathcal{P}}_k(\bar{\bm{H}}_{k}, \bar{\bm{x}}^\text{dig}_k, \bar{\bm{a}}^\text{load}_k),
\end{align}
where $\bar{\bm{x}}_k$ and $\bar{\bm{a}}_k$ are \emph{greedy} choices given the pile state $\bar{\bm{H}}_{k}$, that is, those that minimize $\bm{w}^\mathsf{T}\bm{\mathcal{P}}_k(\bar{\bm{H}}_{k}, \bar{\bm{x}}_k, \bar{\bm{a}}_k)$.
The pile $\bar{\bm{H}}_{k}$ is, in turn, the result of an expansion from the pile state at the previous search depth, $\bar{\bm{H}}_{k-1}$, using a greedy choice at that level too\footnote{If there is a maximum number loading cycles, $N$, then the summation in Eq.~(\ref{eq:Q}) is limited by this in case $n+d > N$.  Also, if $d =1$, there is no second or higher term on the right-hand side.}.
This approximation is to avoid doing an exhaustive search over all action candidates.
Evaluating Eq.~(\ref{eq:Q}) requires that pile state at the top level pile $\bm{H}_n$ is expanded into all the candidate piles $\{\bm{H}_i^{\prime}\}_{i=1}^I$ using $\{\bm{x}_i^\text{dig}\}_{i=1}^I$ and  $\{\bm{a}_i^\text{load}\}_{i=1}^I$.
The algorithm is displayed in Alg.~\ref{alg:search_algorithm} and pictorially illustrated in Fig.~\ref{fig:overview}.

\begin{algorithm} [!htb]
    \caption{Look-ahead tree search with depth $d$ on a planning horizon $N$}
    \label{alg:search_algorithm}
    \begin{algorithmic}
        \State Input: $\{ \bm{H}_1, \bm{x}^\text{dump}, N, d, \bm{w} \}$
        \For{$n \gets 1, \hdots, N$}
            \State $\{ \tilde{\bm{x}}^\text{dig}_i \}_{i=1}^{I} := \texttt{listup}(\bm{H}_n, \bm{x}^\text{dump})$
            \State $\{ \tilde{\bm{a}}^\text{load*}_i \}_{i=1}^{I} := \{ \argmin_{\bm{a}^\text{load}_n} {\bm{w}}^\mathsf{T}\bm{\mathcal{P}}_n^\text{load} \}_{i=1}^{I}$  
            \State $ \bm{a}^\text{load}_n, \bm{x}^\text{dig}_n := \argmin_{\tilde{\bm{a}}^\text{load*}_i, \tilde{\bm{x}}^\text{dig}_i} Q^d_n $ \Comment{Eq.~(\ref{eq:Q})} 
            \State $\bm{\mathcal{P}}_n := \bm{\Psi}(\bm{H}_n, \bm{x}^\text{dig}_n, \bm{a}^\text{load}_n)$
            \State $\bm{H}_{n+1} := \bm{\Phi}(\bm{H}_n, \bm{x}^\text{dig}_n, \bm{a}^\text{load}_n)$
        \EndFor
        \State Output: $\{ \bm{x}^\text{dig}_n, \bm{a}^\text{load}_n, \bm{H}_n, \bm{\mathcal{P}}_n \}_{n=1}^N $
    \end{algorithmic}
\end{algorithm}

\subsubsection{Greedy strategy} 
\label{sec:greedy_strategy}
We refer to the choice of $d=1$ as the entirely \emph{greedy strategy}. The dig location is always selected
to optimize the one-step evaluation function $Q^{d=1}_n = Q_\text{greedy} \equiv \bm{w}^\mathsf{T} \bm{\mathcal{P}_n}$
with the loading action optimized for the local pile state at the dig location.
In this case, the pile state $\bm{H}_n$ needs only to be expanded once and not into multiple pile state candidates.

\subsubsection{Maximum loading strategy} 
\label{sec:maximum_loading_strategy}
Setting $Q^{d=1}_n = Q_\text{loading} \equiv \bm{w}^\mathsf{T} \bm{\mathcal{P}_n}^\text{load}$
results in dig locations and loading actions that are optimized for short-term 
loading performance while ignoring the V-turn cost. We refer to this as the
\emph{maximum loading strategy}.

\subsubsection{Nominal strategy} 
\label{sec:nominal_strategy}
For comparison, we also test the strategy of always selecting the dig location that
has the lowest transportation cost, irrespective of the loading performance.
This is accomplished with  $Q^{d=1}_n = Q_\text{nominal} \equiv \bm{w}^\mathsf{T} \left( \bm{\mathcal{P}_n}^\text{V1} + \bm{\mathcal{P}_n}^\text{V2} \right)$.
We refer to this as the \emph{nominal strategy} as it is a standard method \cite{hong2018path} and the natural choice when there is no access to a world model.

\subsection{Computational time} 
\label{sec:prediction_summary}
The computational time is what limits us from computing the optimal solution by exhaustive search and instead approximate the evaluation function by Eq.~(\ref{eq:Q}) at higher search depths.
We register the computational time for analyzing a single loading cycle. It includes predicting the new pile state $\bm{H}'$ and total performance $\bm{\mathcal{P}}$, which involves updating $\bm{a}^\text{load}$ accorging to Eq.~(\ref{eq:gradient_descent}) (dominant part) and computing and integrating the V-turns given a selected dig location  $\bm{x}^\text{dig}$ as described in Sec.~\ref{sec:vturn_prediction}. Table \ref{tab:sequential_total_outcomes} shows the average computational time from profiling the loading predictions made on a desktop computer with an Intel(R) Core(TM) i7-8700K, 3.70 GHz, 32 GB RAM on a Windows 64-bit system and NVIDIA GeForce RTX 2070 SUPER.
\begin{table} [!htb]
    \centering
    \caption{Profiling a single loading cycle prediction}
    \begin{tabular}{llrl} \toprule
        \multicolumn{2}{c}{Function} & \multicolumn{1}{c}{Time [ms]} & Dominant computing cost  \\ \toprule
        $\bm{\Phi}$ & \texttt{cutout}        & 9.0 & heightfield rotation \\
                            & $\phi$                 & 2.5 &  \\
                            & \texttt{replace}       & 12.0 & heightfield rotation\\ \midrule
        $\bm{\Psi}$ & $\psi^\text{load} \to \bm{a}^\text{load}$   & $\sim45.0$ & \texttt{grad-descent} (1.5 [ms] $\times$ iteration) \\
                            & $\psi^\text{V1}$ & 2.5 & generating 2 $\times$ cubic spline \\
                            & $\psi^\text{V2}$ & 2.5 & generating 2 $\times$ cubic spline \\ \midrule
        \multicolumn{2}{c}{Total} & $\sim73.5$ &  \\ \bottomrule
    \end{tabular}
    \label{tab:sequential_total_outcomes}
\end{table}

The computational time for solving the end-to-end optimization problem depends on the given time for search, the number of child nodes per node ($I$), and the tree search depth ($d$).
For reference, a greedy search takes about 29.4 s for $d=40$, $I=10$ with 73.5 ms per prediction as it requires $I \times d$ predictions. 

\section{Results} 
We test the look-ahead search algorithm considering a wheel loader performing a task of $N=15$ sequential loading cycles.
First, we analyze the greedy strategy (prediction horizon $d=1$) and compare it using the load optimizing and the nominal strategies.
Next, we analyze the look-ahead tree search algorithm with different search depths to examine the effect of the prediction horizon.
The weights were fixed at $ \bm{w}= [2,1,1]$ such that production (loaded mass) and cost (loading time, mechanical work) contribute equally to the overall loading performance.

\begin{figure}[!htb]
    \centering
    \includegraphics[width=0.8\textwidth, trim={0 15 0 0},clip]{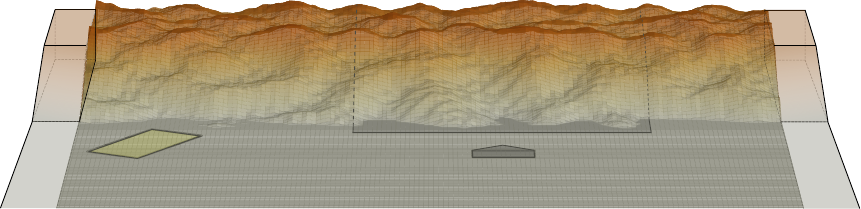}
    \caption{The initial heightfield with the dump truck area indicated to the left and the search region for dig candidates to the right. The size of the bucket is shown in front of the search region.}
    \label{fig:initial_pile_state}
\end{figure}

The initial pile surface $\bm{H}_1$ features a trapezoidal prism, 1.8 m tall with a $30^\circ$ slope at the front, with Perlin noise \cite{perlin1985image} added to the surface to avoid ending up in the same local minima.
A representative initial pile surface is illustrated in Fig.~\ref{fig:initial_pile_state}.
The location and orientation of the receiver are fixed at $\bm{x}^\text{dump} = [x, y, \theta] = [-12.0\text{ m}, -3.0\text{ m}, -30.0^\text{o}]$.
The dig locations are constrained within $-5.0\text{ m} \leq x \leq 8.0\text{ m}$ and $0.0\text{ m} \leq y \leq 6.0\text{ m}$.
The \texttt{listup} function generates candidate dig locations $\tilde{\bm{x}}^\text{dig}$ with a spacing of 1\,m.

\subsection{Greedy strategy} 
\label{sec:first_results}
\begin{figure}[!htb]
    \centering
    \begin{subfigure}[b]{0.6\textwidth}
        \centering
        \includegraphics[width=1.0\textwidth, trim={0 35 0 0},clip]{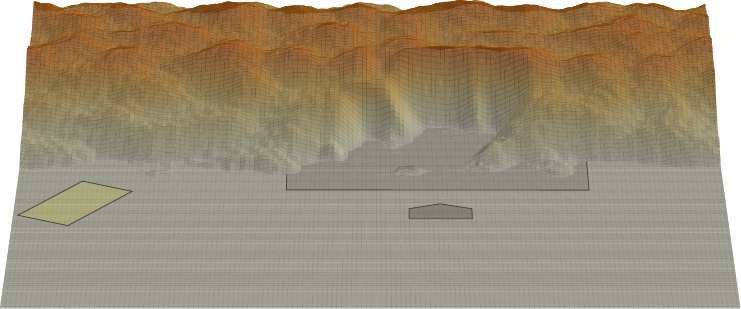}
        \caption{Greedy strategy.}
    \end{subfigure} \\
    \begin{subfigure}[b]{0.6\textwidth}
        \centering
        \includegraphics[width=1.0\textwidth, trim={0 35 0 0},clip]{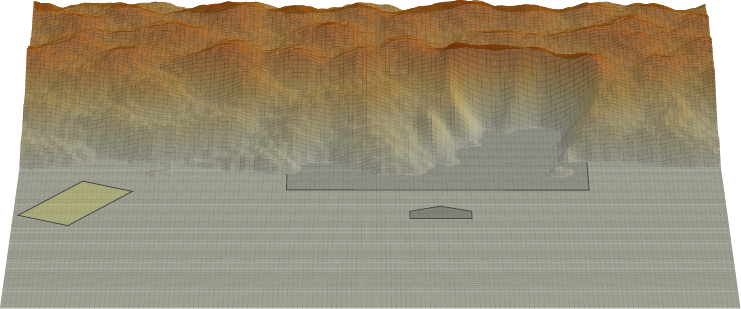}
        \caption{Maximum loading strategy.}
    \end{subfigure} \\
    \begin{subfigure}[b]{0.6\textwidth}
        \centering
        \includegraphics[width=1.0\textwidth, trim={0 35 0 0},clip]{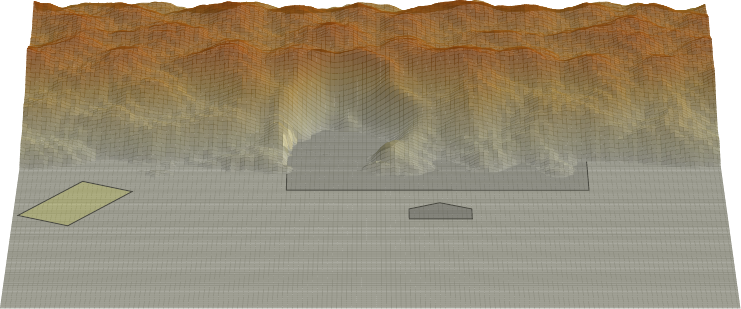}
        \caption{Nominal strategy.}
    \end{subfigure}
    \caption{The pile shape after the five loading sequences using a greedy strategy ($d=1$) and the three different evaluation functions.}
    \label{fig:seq_opt_pile_state_at_5}
\end{figure}

\begin{figure} [!htb]
    \centering
    \includegraphics[width=1.0\textwidth]{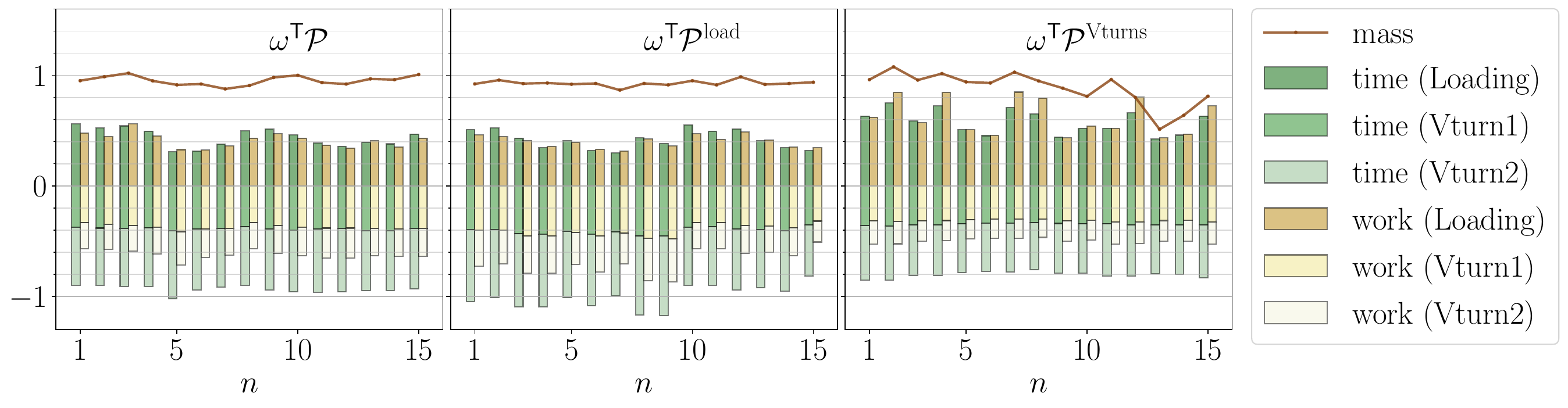}
    \caption{The performance at each loading cycle using the greedy, maximum loading, and nominal strategy. The time and work are split per subtask, normalized, and  V-turn values are negated for ease of comparison.}
    \label{fig:opt_seq_greedy}
\end{figure}

The greedy strategy with no horizon, $d=1$, is tested using the three evaluation functions described in Sec.~\ref{sec:greedy_strategy}-\ref{sec:nominal_strategy}: $Q_\text{greedy}=\bm{w}^\mathsf{T} \bm{\mathcal{P}}_n$, $Q_\text{loading}=\bm{w}^\mathsf{T} \bm{\mathcal{P}}_n^\text{load}$, and $Q_\text{nominal}=\bm{w}^\mathsf{T} \left( \bm{\mathcal{P}}_n^\text{V1} + \bm{\mathcal{P}_n}^\text{V2} \right)$.
The pile states after five loading cycles are shown in Fig.~\ref{fig:seq_opt_pile_state_at_5}.
The greedy strategy results in loadings that aim for areas in the search region with more soil easily accessible from the dump truck. The maximum loading strategy aims to the right where there is more soil but further from the dump truck.
The nominal strategy loads to the left, in proximity to the dump truck, as can be expected when focusing entirely on minimizing the transportation cost. 
The evolution of the loaded mass, time, and work over 15 loading cycles is shown in Fig.~\ref{fig:opt_seq_greedy}, and the net result is listed in Table~\ref{tab:opt_seq_greedy_sum}. The greedy strategy performs the best on average and in total. The nominal strategy's performance drops over time, suggesting that the pile state is deteriorating.
The greedy strategy is 8\,\% more productive and energy efficient than the nominal strategy, using nearly the same time to complete 15 cycles. The maximum loading strategy is  1\,\% more productive than the greedy strategy in loading tasks but spends around 8\,\% more time and energy during V-turns. That is an expected result, as transportation time and energy do not contribute to the evaluation function of the maximum loading strategy.
The analysis demonstrates that the evaluation function works effectively.
\begin{table} [!htb]
    \centering
    \caption{The total performance by the three variants of the greedy search strategy in units $M$ [tonne], $T$ [s], and $W$ [MJ].}
    \label{tab:opt_seq_greedy_sum}
    \scalebox{0.7}{
        \begin{tabular}{l c c c c c c c}
        \dtoprule
        &  & \multicolumn{2}{c}{Load} & \multicolumn{2}{c}{V-turns} & \multicolumn{2}{c}{Total} \\
        \cmidrule(r){3-4} \cmidrule(r){5-6} \cmidrule(r){7-8}
        Strategy & $M$ & $T$ & $W$ & $T$ & $W$ & $T$ & $W$ \\ \dtoprule
        Greedy & $\bm{64.4}$ & 197 & 6.2 & 421 & 9.3 & $\bm{694}$ & $\bm{15.5}$ \\
        Max loading & 62.6 & $\bm{189}$ & $\bm{6.0}$ & 453 & 10.4 & 717 & 16.4 \\
        Nominal & 59.8 & 260 & 9.4 & $\bm{362}$ & $\bm{7.5}$ & 697 & 16.9 \\
        \dbottomrule
        \end{tabular}
    }
\end{table}

\subsection{Long-horizon planning with look-ahead tree search} 
\label{sec:second_results}
\begin{figure} [!htb]
    \centering
    \includegraphics[width=0.5\textwidth, trim = 0 0 0 0, clip]{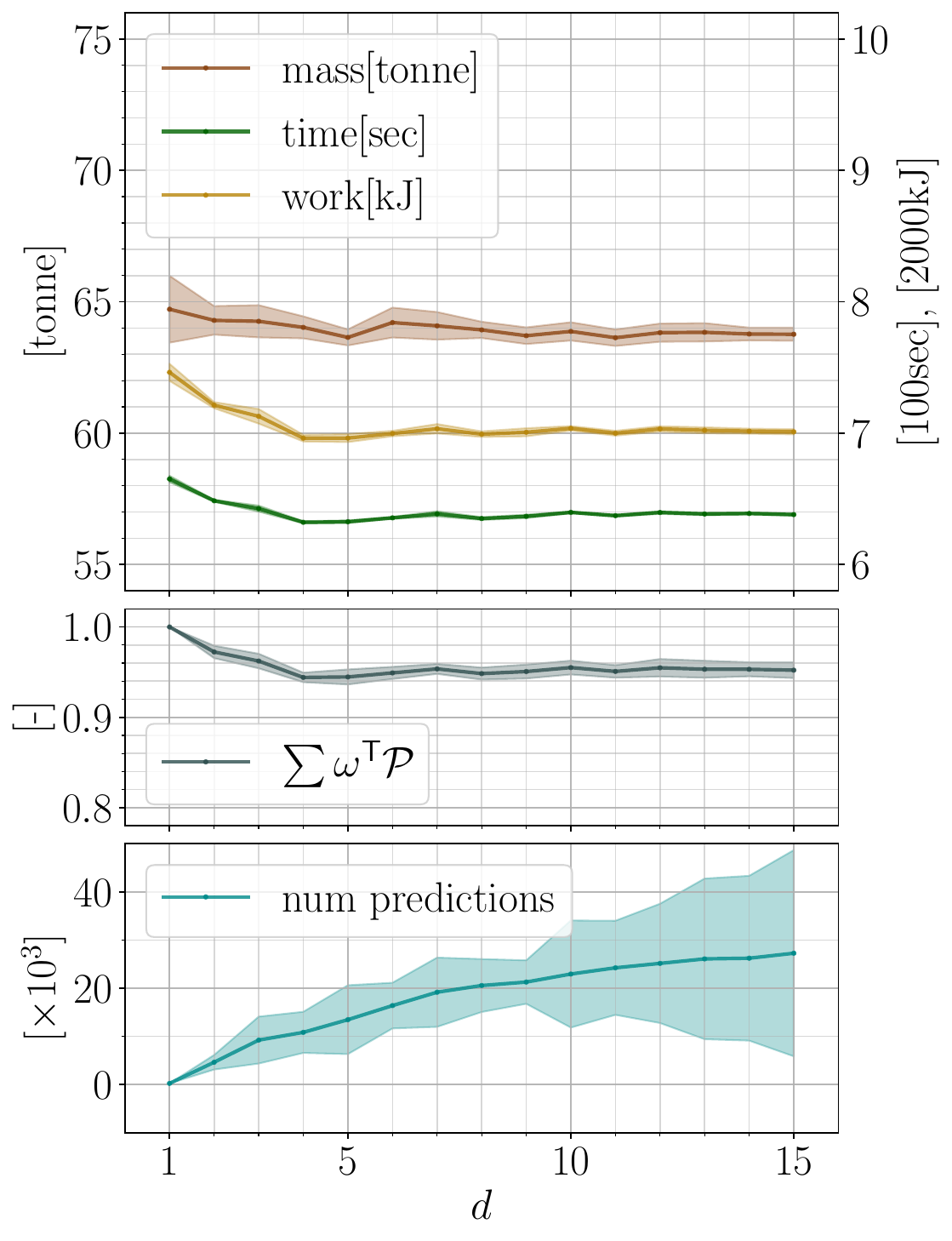}
    \caption{The trend of the total performance of the optimal sequence, evaluated values, and the total number of the predictions at each search for the prediction horizon $d$.
    The values are the averages of all test results with the ten different initial piles.}
    \label{fig:comparison_roll_out_depth}
\end{figure}

The second test uses the tree search algorithm with finite prediction horizon $d$ and evaluation function~(\ref{eq:Q}).
Ten initial heightfields were prepared with different Perlin noise applied to the 1.8 m high trapezoidal prism. 
For each of these ten piles, $N=15$ sequential loading cycles were evaluated with the prediction horizon ranging from $d=1$ to $d=N$.
The resulting loading outcome and performance evaluation function are shown
in Fig.~\ref{fig:comparison_roll_out_depth}. The results have been averaged over the ten repetitions starting with the different initial piles.
The performance converges around $d=4$. Beyond this point, increasing the search horizon further does not enhance the overall performance.

The performance improves by about 5.6\,\% on average when comparing the results between $d=1$ and $d=4$. The total loaded mass is marginally reduced from about 64.7 to 64.0 tonnes (1.1\,\%), while the total loading time and the work improves from about 665 to 632 s (5.0\,\%) and from 14.9 to 13.9 MJ (6.7\,\%), respectively.
Note that we also tested with evaluation horizons of $N=10$ and $N=5$.
The results at $N=10$ showed a similar improvement ratio, but this trend was not found for $N=5$, probably because of the shorter task length.
As we have already found that the greedy strategy ($d=1$) is 6\,\% more performant than the nominal strategy, we conclude that the look-ahead tree search leads to a 14\,\% increase in performance relative to the nominal strategy.

\begin{figure} [!htb]
    \centering
    \includegraphics[width=0.5\textwidth, trim = 0 0 0 0, clip]
    {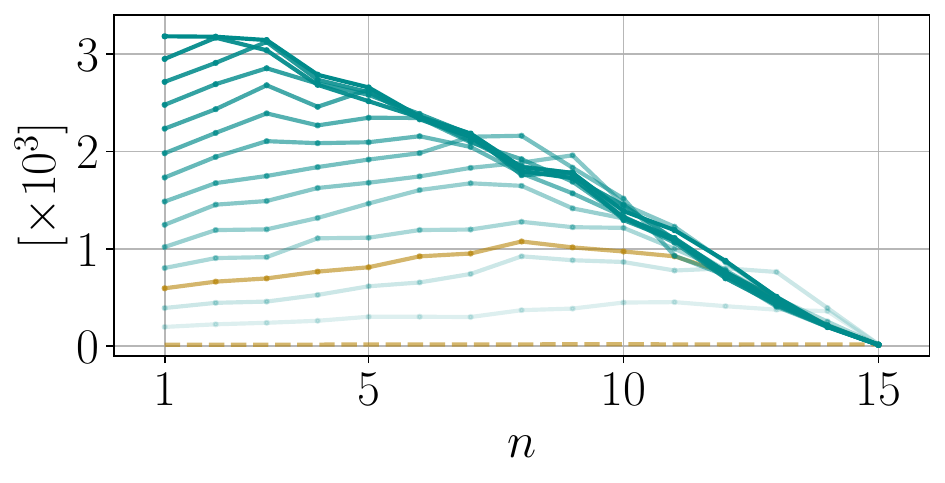}
    \caption{The number of the predictions per loading cycle in a search with planning horizon $N=15$.  The line transparency increases with prediction horizon $d$ from $d=2$ (light) to $d=15$ (dark). The golden lines are the trend of $d=1$ (dashed) and $d=4$ (solid).}
    \label{fig:comparison_roll_out_depth_per_cycle}
\end{figure}

The computational cost of the search increases with prediction horizon $d$.
On average, the search took 250 and 10,843 predictions with $d=1$ and $4$, respectively.
The computational time was about 18 and 792 s, respectively, given that a prediction takes 73 ms on average.
The computational cost for deciding an immediate optimal action is shown in Fig.~\ref{fig:comparison_roll_out_depth_per_cycle}.
The search with $d=15$ required 3,181 predictions to decide the first action, which amounts to 232 s, given a prediction takes 73 ms on average.
The searches with $d=4$ and $d=1$
required 593 and 13 predictions to decide the first action, amounting to 43 s
and 0.9 s, respectively.
Note that the number of predictions depends on the prediction horizon, the resolution of \texttt{listup}, and the length along the pile edge within the constraint region.
The loading action sequence can make the pile state complex, which can increase the number of predictions.

\begin{figure} [!htb]
    \centering
    \begin{subfigure}[b]{0.60\textwidth}
        \centering
        \includegraphics[width=1.0\textwidth, trim = 0 0 0 0, clip]{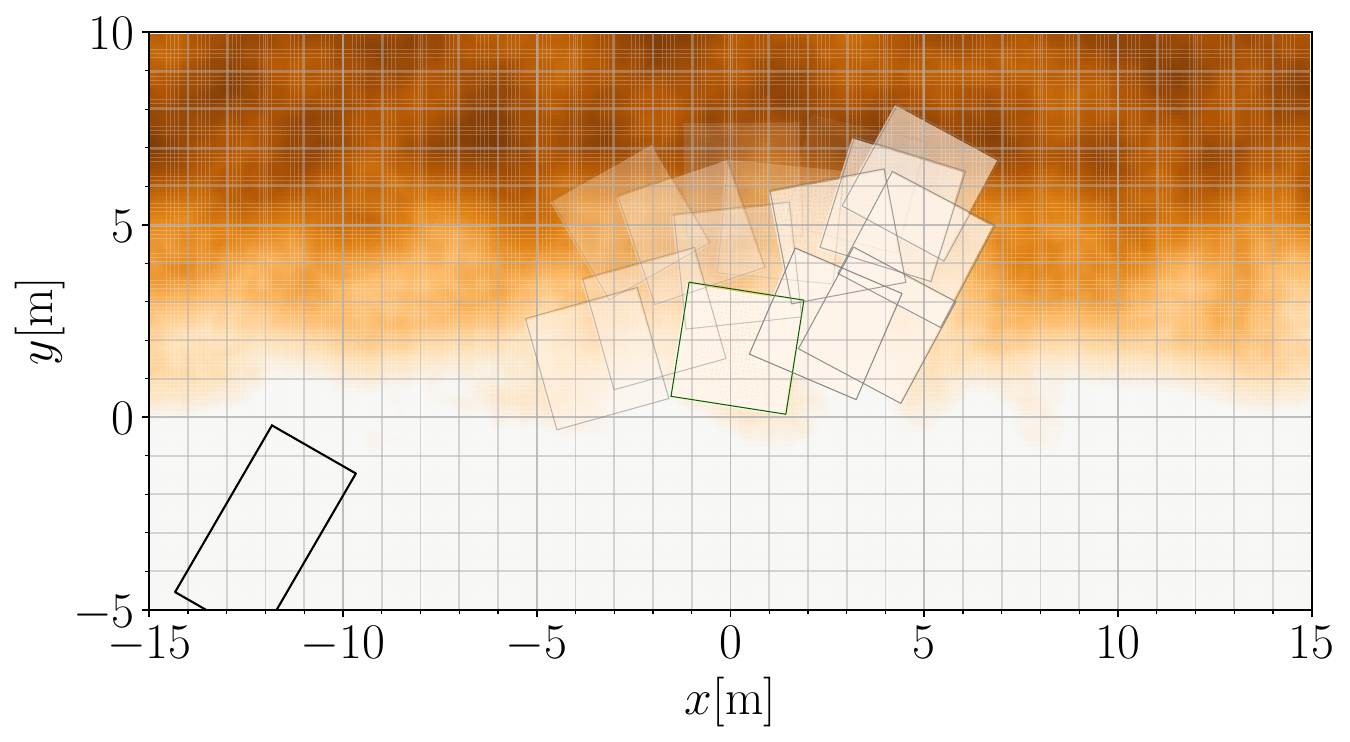}
        \caption{$d=1$}
        \label{fig:opt_sequence_a}
    \end{subfigure} \\
    \begin{subfigure}[b]{0.60\textwidth}
        \centering
        \includegraphics[width=1.0\textwidth, trim = 0 0 0 0, clip]{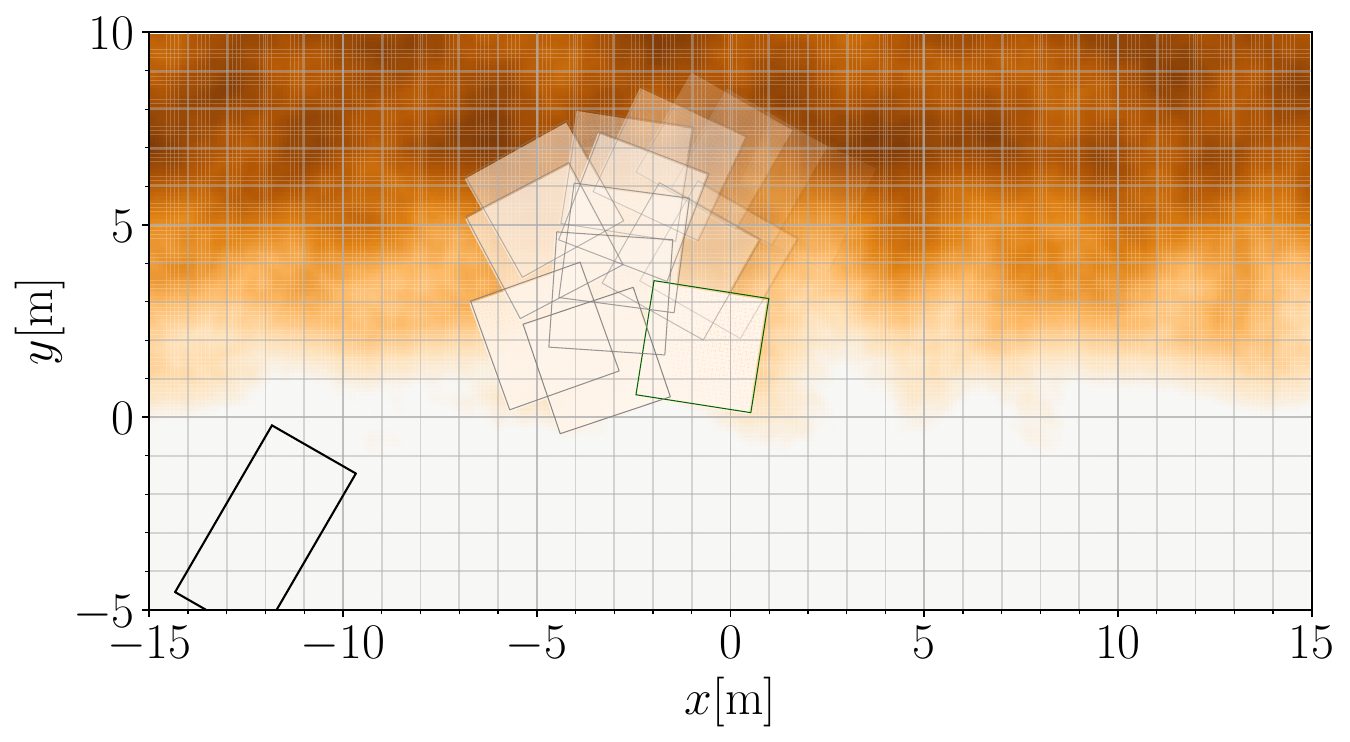}
        \caption{$d=4$}
        \label{fig:opt_sequence_b}
    \end{subfigure}
    \caption{Visualization of the optimized sequence of dig locations $\{ \bm{x}_n^\text{dig} \}_{n=1}^15$ computed using greedy search ($d=1$) and tree search (depth $d=4$).
        The first dig location is marked with a green frame, and later locations with increasing transparency.}
    \label{fig:opt_sequence}
\end{figure}

The optimal loading locations are visualized in Fig.~\ref{fig:opt_sequence} for prediction horizons $d=1$ and $4$.
The first dig locations are rather similar, but the consecutive sequences are distinct from each other.
The greedy search in Fig.~\ref{fig:opt_sequence_a} starts at $x=0$, continues to the right, and then switches to the left.
The tree search method in Fig.~\ref{fig:opt_sequence_b} starts a little closer to the dump truck and stays consistently to the left, closer to the dump truck, working its way into the pile with approximately three buckets' width.
The conclusion is that the tree search method transforms the pile in a way that maintains future loading with good outcomes while keeping proximity to the dump truck.

\section{Discussion} 
With the look-ahead search method, it is possible to find action sequences that maximize the total performance over long horizons by increasing the pile state quality for future loadings. This method adapts to the truck's location and can do so even if it would vary provided that the location is known in advance.
In our experiments, the optimal action sequence can be (before any serious code optimization) computed in 43 s with a look-ahead search with depth $d = 4$, which is close to the average loading cycle time.
Future work should consider other soil properties, such as cohesive or inhomogeneous soil. The resulting pile state is likely to depend more strongly on the selected loading actions than it does for gravel, for which the pile always tends to settle with a local slope near the natural angle of repose. 
To support more realistic conditions, one should extend the framework to support non-flat ground and spillage.
Capturing this with a world model would be more demanding and might benefit from a longer horizon than $d = 4$. That would motivate substantial improvements in the computational efficiency of the look-ahead search.
In this study, we employed a set of specific weight factors that equally balance productivity (mass per unit time) and energy efficiency (mass per unit energy). It would be interesting, in future work, to explore the impact of different weight factors on the result, e.g., what digging behaviour emerges when energy consumption is heavily penalized.



\section{Conclusion} 
We describe and test a look-ahead tree search method to find near-optimal loading action sequences. This method uses a world model to predict future pile states and the performance for each loading cycle. It can dynamically adjust to varying conditions and constraints of the work site. 
Our results show significant performance improvements, showcasing the benefits of incorporating long-term predictions into decision-making.
The look-ahead tree search results in 6\,\% higher performance in sequential loading than a greedy strategy and 14\,\% higher than the nominal strategy.
Moreover, the tree search using long-horizon predictions leads to different decisions for immediate actions.
This raises the question for future work on how sensitive the optimized action sequence is to variations in soil properties.

\section*{Acknowledgement}
The research was supported in part by Komatsu Ltd and Algoryx Simulation AB.
We thank Erik Wallin and Arvid Fälldin for providing us the valuable suggestions and implementation to improve the prediction speed.

\section{References}
\printbibliography

\end{document}